\documentclass[aps,reprint,prl,superscriptaddress]{revtex4-1}
  \usepackage{ulem}
  \usepackage{graphicx}
\usepackage{xcolor}
\usepackage{hyperref} 
\usepackage{subfigure}
\usepackage{multirow}

  \def\CTEP{compound-tunable embedding potential}
  \def\abinitio{{\it ab~initio}}
  
  \def\dfel{$d/f-$}

\begin{document}
\title{Compound-tunable embedding potential method and its application to ytterbium fluoride crystals YbF$_2$ and YbF$_3$}
\author{V.M.\ Shakhova}\email[]{verahcnkrf@gmail.com}
\affiliation{Petersburg Nuclear Physics Institute named by B.P.\ Konstantinov of National Research Center ``Kurchatov Institute'' (NRC ``Kurchatov Institute'' - PNPI), 188300, Russian Federation, Leningradskaya oblast, Gatchina, mkr.\ Orlova roscha, 1}
\affiliation{Saint Petersburg State University, 7/9 Universitetskaya nab., 199034 St. Petersburg,  Russia}
\author{D.A.\ Maltsev}
\affiliation{Petersburg Nuclear Physics Institute named by B.P.\ Konstantinov of National Research Center ``Kurchatov Institute'' (NRC ``Kurchatov Institute'' - PNPI), 188300, Russian Federation, Leningradskaya oblast, Gatchina, mkr.\ Orlova roscha, 1}
\author{Yu.V.\ Lomachuk}
\affiliation{Petersburg Nuclear Physics Institute named by B.P.\ Konstantinov of National Research Center ``Kurchatov Institute'' (NRC ``Kurchatov Institute'' - PNPI), 188300, Russian Federation, Leningradskaya oblast, Gatchina, mkr.\ Orlova roscha, 1}
\author{N.S.\ Mosyagin}
\affiliation{Petersburg Nuclear Physics Institute named by B.P.\ Konstantinov of National Research Center ``Kurchatov Institute'' (NRC ``Kurchatov Institute'' - PNPI), 188300, Russian Federation, Leningradskaya oblast, Gatchina, mkr.\ Orlova roscha, 1}
\author{L.V.\ Skripnikov}
\affiliation{Petersburg Nuclear Physics Institute named by B.P.\ Konstantinov of National Research Center ``Kurchatov Institute'' (NRC ``Kurchatov Institute'' - PNPI), 188300, Russian Federation, Leningradskaya oblast, Gatchina, mkr.\ Orlova roscha, 1}
\affiliation{Saint Petersburg State University, 7/9 Universitetskaya nab., 199034 St. Petersburg,  Russia}
\author{A.V.\ Titov}\email[]{titov\_av@pnpi.nrcki.ru}
\affiliation{Petersburg Nuclear Physics Institute named by B.P.\ Konstantinov of National Research Center ``Kurchatov Institute'' (NRC ``Kurchatov Institute'' - PNPI), 188300, Russian Federation, Leningradskaya oblast, Gatchina, mkr.\ Orlova roscha, 1}
\date{\today, version: `EP-YbFn-11sh'}
\begin{abstract}
 Compound-tunable embedding potential (CTEP) method developed in \cite{Maltsev:19a, Lomachuk:19a} to describe electronic structure of fragments in materials is applied to crystals containing periodically arranged lanthanide atoms, which can have open $4f$ shell. We consider YbF$_2$ and YbF$_3$ as examples such that $4f$ shell is excluded from both the crystal and cluster stages of generating the CTEP. Instead, 10 and 11 valence-electron pseudopotentials for Yb, correspondingly, are applied and the latter treats the $4f$-hole implicitly. At the next stage of the two-component embedded cluster studies of the YbF$_{2,3}$ crystals we apply the 42 valence-electron relativistic pseudopotential for Yb and, thus, $4f$ shell is treated explicitly. A remarkable agreement of the electronic density and interatomic distances within the fragment with those of the original periodic crystal calculation is attained. 
\end{abstract}
 
\maketitle

\section{Introduction}
\label{Intro}
At present, large diversity of materials with various exceptional properties provides wide range of both unique and multifunctional applications. 
There are the materials which contain transition metals ($d-$elements), lanthanides, and actinides ($f$- and \dfel elements) \cite{Khomskii:14_d-el, Martin-Ramos:18_Ln, Cotton:06_An}, either as regular atoms in periodic structures or impurities. Such materials are most challenging for their theoretical study. 
Ytterbium trifluoride, YbF$_3$, one of such materials is widely used in fluoride glasses. For example, using a coating by YbF$_3$ film over the Eu-doped SnO$_2$ crystal, it is possible to modify its characteristics. 
This coating plays a significant role in development of solar cells \cite{yue:17}. 
In turn, one can improve characteristics of some other crystals by doping them with YbF$_3$: the fluoride-based ceramics such as CaF$_2$ are known as excellent optical conductors used in diode laser pumping \cite{carnall:66, vydrik:80,siebold:09}, however, doping the ceramics by YbF$_3$ (with minimal 3\% admixture) leads to superior mechanical properties of the ceramics \cite{akchurin:13}. 
The addition of YbF$_3$ to calcium silicate cements leads to such changes in properties of the material that it becomes more suitable for dental applications \cite{antonijevic:15}.

Also new era in investigating  materials and defects containing heavy transition metals, lanthanides and actinides was opened due to impressive recent achievements in creating experimental facilities like X-ray free-electron lasers, synchrotrons \cite{Zubavichus:01rev, Chergui:16_StrDin}, etc., to study local atomic-scale electronic structures in materials.

However, theoretical possibilities in direct studying of electronic structures at atomic-scale and different physical-chemical properties including those concentrated on heavy atoms
\cite{Titov:14a, Skripnikov:15b, Lomachuk:13} are yet hampered by several challenges in quantum chemical description of such systems.
At first they require a simultaneous treatment of relativistic and correlation effects at a very high level.
Besides, polyvalent \dfel elements often have pronounced multireference character and high density of low-lying electronic states. 
As a result, opportunities for direct \abinitio\ study of materials containing \dfel elements with required accuracy can be blocked by unacceptable computational costs (see analysis for ThO in \cite{Skripnikov:15a}).
An alternative way to explore such a material is to reduce its studying to a molecular-type problem for some of its fragment, assuming that relaxation of the rest of the crystal (environment of the fragment) in processes under consideration is negligible. 
In this case one can consider influence of the environment on the fragment by some approximate {\it embedding potential} to improve the quality of description of phenomena localized on the fragment using extended possibilities of its studying by molecular methods. 
Such a fragment with embedding potential is usually called the ``embedded cluster'' or ``cluster, embedded in a crystal''.

This work is devoted to theoretical modeling of ytterbium di- and tri-fluorides using compound-tunable embedding potential method described in work \cite{Maltsev:19a, Lomachuk:19a}.

\section{Method}

The two-component DFT method \cite{Wullen:10} with hybrid PBE0 \cite{Adamo:99} functional implemented in the {\sc crystal} code \cite{Dovesi:18} was used to carry out calculations of the electronic structure of the YbF$_{2}$ and YbF$_{3}$ crystals.
This code is well suited to study ionic-covalent structures, it allows us to use the same DFT PBE0 functional and basis sets, which are used in the cluster model calculations (see below). 
So, one can directly juxtapose results of these calculations.  
For simulation of the crystal structure, the basis set overflow arising in presence of diffuse orbitals can be significant, since relatively small atomic basis sets are used for such studies to avoid their linear dependence. 
In turn, they can be well corrected in cluster calculations with {\CTEP} (CTEP) \cite{Lomachuk:19a, Maltsev:19a}.
The electronic properties of the Yb atom in YbF$_2$ and YbF$_3$ crystals are studied using the following cluster model.
The ytterbium atom is selected as the central atom of the cluster and is described using pseudopotentials with the large number of valence and subvalence electrons (explicitly accounted for in the calculations) and the corresponding basis sets. Its immediate environment consists of fluorine atoms, for which all electrons are treated explicitly with corresponding basis sets. 
The combination of these Yb and F atoms will be called as the main-cluster. 
The pseudopotentials for Yb and basis sets for all main-cluster atoms used in the cluster calculations are the same as those in the solid-state ones.

Note, that for simulation of a crystal fragment with central Yb and only nearest F atoms, the main cluster should have negative charge to reproduce the oxidation state of these atoms. The main-cluster environment have to compensate it by positive charges composing CTEP (and contributing to the extended cluster model). CTEP consists from the {\it nearest cationic environment} and the {\it nearest anionic environment}.

The {\it nearest cationic environment} (or NCE layer) of the cluster model consists of the ytterbrium cations, which are modeled by particular kind of the ``electron free'' semi-local pseudopotentials for Yb$^{+2}$ and Yb$^{+3}$ (corresponding to its oxidation state +2 and +3 in ytterbium di and trifluorides crystals), positive charges and the corresponding basis sets.
Note, that the set of NCE pseudopotentials (without Coulomb terms from their point charges) constitute the {\it short-range} part of CTEP.

The {\it nearest anionic environment} (or NAE layer) consists of the fluorine anions, which are modeled by just negative point charges without any semilocal part. 

In the present work, values of these charges are obtained by minimizing 
root mean square (RMS)
force $|f|$ acting on the atoms of the main cluster. This 
RMS
force is calculated as 
\begin{equation} 
\label{eq:av_force} 
|f| = \sqrt{\sum\limits_{i=1}^{N_{at}}\left(\nabla_i E)^2\right/N_{at}}
\end{equation} 
$E$ is the total energy of the cluster, $N_{at}$ is the number of atoms in the main cluster ($N_{at}=9$ for YbF$_2$ and $N_{at}=10$ for YbF$_3$), and $\nabla_i$ is the gradient operator with respect to coordinates of $i$-th atom.

\section{Computational details}

All calculations were performed using DFT method with PBE0 functional. The solid-state and cluster calculations were performed with {\sc crystal-17} \cite{crystal:17} and relativistic molecular DFT \cite{Wullen:10} packages, respectively.

The relativistic pseudopotentials (PP) generated by our group \cite{Titov:99} for the Yb atom were used, namely 42ve-PP, 10ve-PP, and 11ve-PP.
The first one includes 28 electrons, 
[Ar]3d$^{10}$, in the pseudopotential, and the remaining 42 electrons, 4s$^2$4p$^6$4d$^{10}$4f$^{14}$5s$^2$5p$^6$6s$^2$, are explicitly taken into account in the calculation. 
Inclusion of the 4$f$ shells is important for accurate calculations of properties localized on an atom. 
However, in this case, the PP has too many explicitly treated electrons, which leads to either a significant increase in the required computer resources or unstable solutions (especially in crystal calculations).

The 4$f$ shell in lanthanides is valence in terms of its energy but is localized in the outer-core spatial region. 
Therefore, in cases where it is known for sure that the occupation number of the 4$f$ shell does not change, one can use 10ve-PP, in which 60 electrons, 
[Kr]4d$^{10}$4f$^{14}$, are in the PP and 10 electrons, 5s$^2$5p$^6$6s$^2$, are outside.
In the cases where the 4$f$ shell (of trivalent Yb) has one electronic hole, we use the special hole-shape PP, or 11ve-PP, generated in this paper. 
The 59 electrons, 
[Kr]4d$^{10}$4f$^{13}$, are excluded from the following calculations with this PP and the 11 electrons, 5s$^2$5p$^6$6s$^2$5d$^1$, are considered 
(PP was built for the generator state when one of the 4$f$-electrons was excited in the 5$d$ shell). 
It is demonstrated below that such PPs work well when crystal geometry is optimized.

The basis sets, corresponding to these PPs are
(8,8,6,7)/[7,6,4,4]
for 42ve-PP,
(6,6,2,2)/[4,4,2,2]
for 10ve-PP,
(7,8,6,7)/[6,6,4,4]
for 11ve-PP.

The basis sets: (4,3,4,1)/[3,2,2,1] or (3,2,3,1)/[3,2,3,1] with the corresponding tuned 0ve-PPs for the Yb ``pseudoatoms'' treated as NCE layer were used in the cases of YbF$_2$ or YbF$_3$.

The fluorine atoms are treated at all-electron level, basis sets for them are taken from \cite{Peintinger:13}.

\section{Results and discussions}

\textbf{1. Periodic structure calculations}

The ytterbium(II) fluoride crystal belongs to the {\it Fm-3m} space group and consist of two non-equivalent atomic types: Yb and F \cite{Kompanichenko:10}. 
The ytterbium(III) fluoride crystal belongs to {\it Pnma} space group and consist of three non-equivalent atomic types: Yb and two different F types \cite{Bukvetskii:77}. 
Both atomic positions and cell parameters were optimized with only crystal symmetry group fixed (see Table \ref{table:crystall}). 
\begin{table}[h!]
	\caption{Structural parameters of ytterbium di- and trifluoride crystals (in {{\AA}})}
	\label{table:crystall} 
	\begin{tabular}{ccccc}
		\hline
		\hline
		Crystal & Cell parametres & Experiment & 10(11)ve-PP & \% \\
		\hline
		\hline 
		YbF$_2$ {\it Fm-3m} & a=b=c & 5.608 \cite{Kompanichenko:10} & 5.532 & 1.3 \\ 
		\hline 
		& a & 6.219 \cite{Bukvetskii:77} & 6.185 & 0.5 \\
		YbF$_3$ {\it Pnma} & b & 6.787 \cite{Bukvetskii:77} & 6.747 & 0.6 \\
		& c & 4.434 \cite{Bukvetskii:77} & 4.385 & 1.1 \\ 
		\hline
		\hline
	\end{tabular}
\end{table}
The YbF$_2$ crystal structure obtained using 10veP was close to the experimental one with the cell parameter error about 1.3\%.
The YbF$_3$ crystal structure obtained using 11veP was close to the experimental one with cell parameter error of 0.5-1.1\%.

As one can see from Table \ref{table:crystall}, parameters of the unit cells of crystals YbF$_2$ and YbF$_3$ obtained using 10ve-PP and 11ve-PP are in good agreement with experimental data \cite{Kompanichenko:10,Bukvetskii:77}.
This suggests that the constructed PPs adequately describe the electron density in these compounds.

Calculations of the crystals periodic structure structures with the use of the 42vePP are diverged.

\textbf{2. Cluster calculations}

\begin{figure}[h]
	\centering
	\begin{minipage}[h]{0.44\linewidth}\centering{\includegraphics[width=\linewidth]{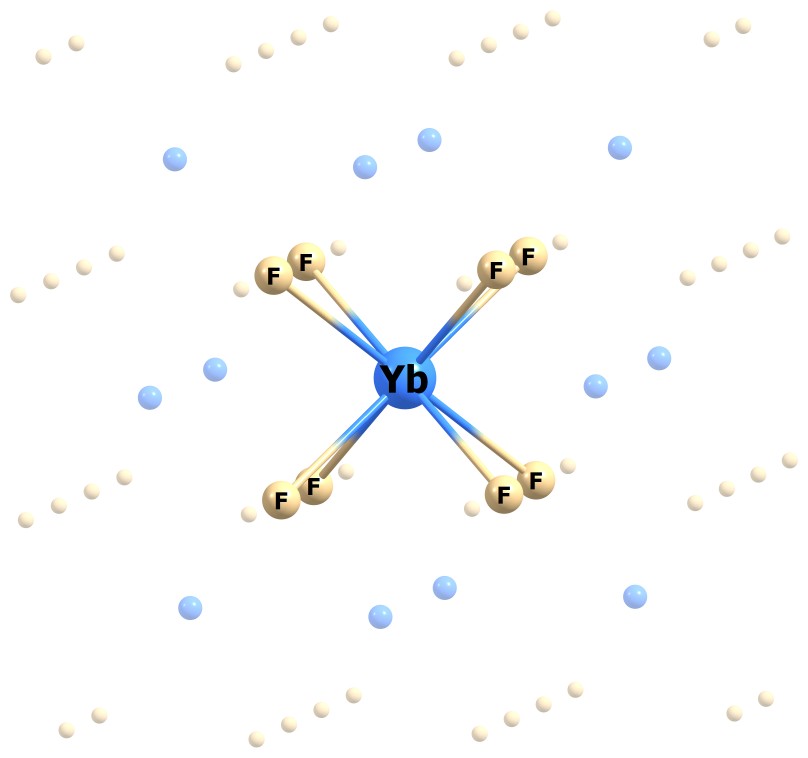} \\ (a)}
	\end{minipage}
	\begin{minipage}[h]{0.54\linewidth}\centering{\includegraphics[width=\linewidth]{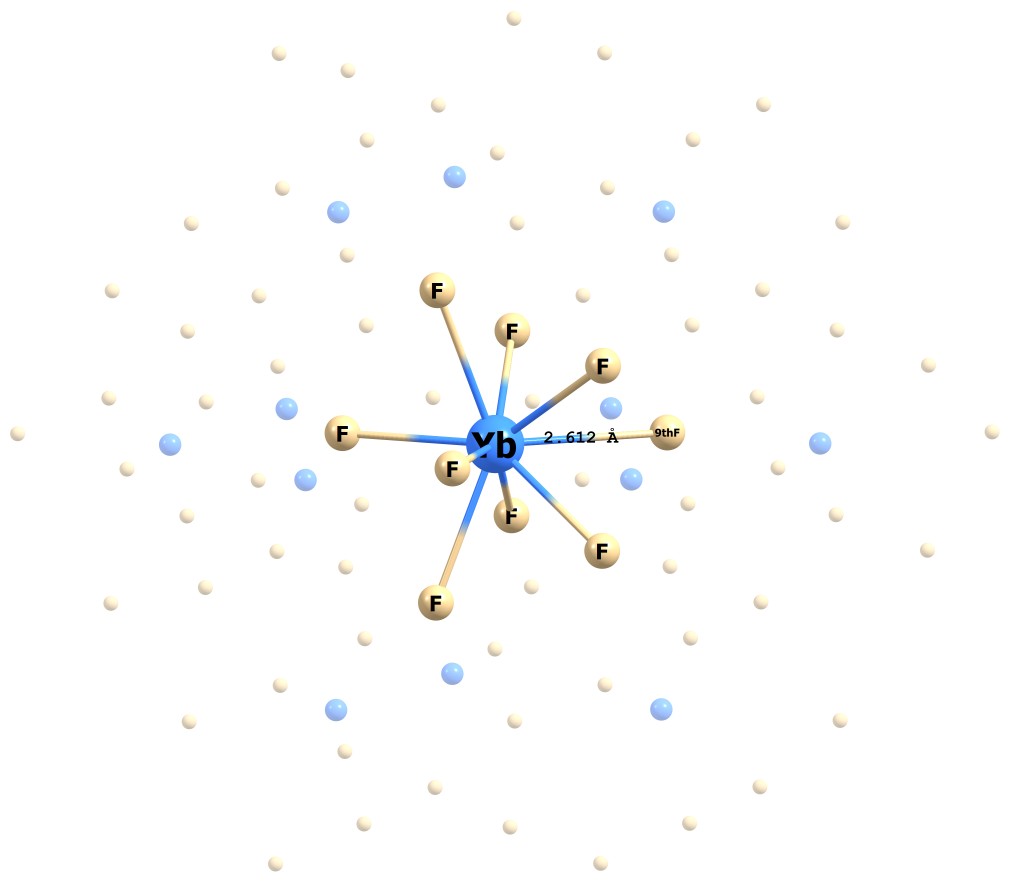} \\ (b)}
	\end{minipage}
	\caption{Yb-centered clusters:  $($a$)$ [YbF$_{8}$][Yb$_{12}$][F$_{48}$], and $($b$)$ [YbF$_{9}$][Yb$_{12}$][F$_{65}$]. 
		NCE and NAE atoms are shown as spheres of half-radius without caption (of the same colour as for corresponding atoms of the main 
		cluster)} 
	\label{fig:cluster_ybf}
\end{figure}

YbF$_2$ is a high symmetry crystal, so that the process of cluster building is straightforward.
The main cluster consists of the Yb atom in the center and 8 F atoms as the nearest ones, which form a cube around (Figure \ref{fig:cluster_ybf} a).
The total formula is [YbF$_{8}$][Yb$_{12}$][F$_{48}$] (designated as YbF$_2$(cl)).

YbF$_3$ is a low symmetry crystal, and there is an uncertainty in building of the cluster model.
Each ytterbium atom in the YbF$_3$ crystal has 8 nearest fluorine atoms located at the distance of 2.222$\div$2.289{{\AA}}, but there is also the ninth F atom at the distance of 2.612{{\AA}} (Figure \ref{fig:cluster_ybf} b). 
It is the nearest F atom to another Yb atom. 
The study of the structure of this crystal was made in works \cite{Garashina:77,Garashina:80}. 
The conclusion that this ninth F atom can not be considered the closest atom for the central Yb atom was done. 
However, this distance is comparable to the sum of the covalent radii of ions (1.16 and 1.19{{\AA}} for Yb$^{+2}$ and F$^-$ respectively \cite{Shannon:76}), which means that this 9th F atom will affect the center Yb atom, so that for building of the Ytterbium-centered cluster we have to include this atom in the first coordination sphere and take into account in calculations explicitly.
Thus, the cluster for YbF$_3$ crystal has a form [YbF$_{9}$][Yb$_{12}$][F$_{65}$] (designated as YbF$_3$(cl)).

For each of the cluster with CTEP, the fractional charges on the NCE and NAE layers were optimized. 
The position of all atoms was taken from crystalline calculations with 10ve-PP and 11ve-PP. 
The resulting RMS forces are 2.2$\cdot$10$^{-5}$ a.u.\ for YbF$_2$(cl) and 2.5$\cdot$10$^{-3}$ a.u.\ for YbF$_3$(cl).

\textbf{3. Cluster optimization}

For a verification, positions of main cluster atoms were optimized under different conditions. 
First, not only 10ve-PP on the ytterbium atom was used, but also 42ve-PP. 
Secondly, the positions of the CTEP atoms were taken both from optimized crystal calculations (CTEP) and experimental data (CTEP-expgeom), but the charges were optimized in both cases.
The resulting displacements are given in Table~\ref{table:distort}.

\begin{table*}[h!]
	\caption{Yb--F bond lengths in the main clusters of YbF$_2$(cl) and YbF$_3$(cl) and averege atomic displacements within the main cluster after its optimizations, {{\AA}}}
	\label{table:distort} 
	\begin{tabular}{ccccc}
		\hline
		\hline
		\multirow{3}{*}{Calculations} & Structure & Displacements & Structure & Displacements \\
		\cline{2-5}
		& YbF$_2$(cl): Yb--F$^{(1)-(8)}$ & $|$Calculations--exp$|$ & YbF$_3$(cl): Yb--F$^{(1)\div(8),(9)}$ & $|$Calculations--exp$|$ \\
		\hline
		\hline
		Experiment & 2.428 \cite{Kompanichenko:10} & 0 & 2.222$\div$2.289, 2.612 \cite{Bukvetskii:77} & 0 \\
		\hline
		Crystal-opt & 2.396 & 0.033 & 2.215$\div$2.273, 2.582 & 0.017 \\
		42ve-PP/CTEP & 2.414 & 0.014 & 2.250$\div$2.297, 2.563 & 0.015 \\
		10(11)ve-PP/CTEP & 2.396 & 0.033 & 2.239$\div$2.295, 2.558 & 0.019 \\
		42ve-PP/CTEP-expgeom & 2.443 & 0.014 & 2.242$\div$2.311, 2.614 & 0.011 \\ 		
		10(11)ve-PP/CTEP-expgeom & 2.424 & 0.004 & 2.230$\div$2.309, 2.611 & 0.011 \\
		\hline		
		\hline
	\end{tabular}
\end{table*}

\section{Conclusions}

Compound-tunable embedding potential method developed by us in \cite{Maltsev:19a, Lomachuk:19a} to describe electronic structure of fragments which may contain defects with $f-$elements is applied to crystals containing periodically arranged lanthanide atoms. 
The YbF$_2$ and YbF$_3$ crystals are considered such that $4f$ shell is excluded from both the crystal and cluster stages of generating the CTEP to make calculations with {\sc crystal} code stable. 
Typical DFT accuracy for such kind of crystals, within 0.03~{{\AA}} deviation from experimental interatomic distances, is attained. 
Thus, 10ve ($5s^2 5p^6 6s^2$) and 11ve ($5s^2 5p^6 6s^2 5d^{1}$) large-core PPs for Yb, correspondingly, are applied describing core  [...] $4f^{14,13}$ shells implicitly. 
At the two-component embedded cluster studies of the YbF$_{2,3}$ crystals we apply the small-core 42ve-RPP for Yb and, thus, all $4f$-electrons are treated explicitly.

Our results show applicability of the CTEP method to study local electronic structure and properties of crystals containing periodically arranged lanthanide atoms in the cases when direct treatment of them with using precise versions of small-core PP is embarrassing due to necessity of treating core electrons with spin-orbit effects taken into account and limitations with the basis set size in studying periodic systems. 
First, as one can see from Table~\ref{table:distort} the Yb-F distances scales properly with changing the positions of CTEP atoms (that means transferability of CTEP, see lines in Table~\ref{table:distort} with CTEP-expgeom, corresponding the experimental position of CTEP atoms).
Second, the replacement of 10(11)ve-PPs by 42ve-PPs leads to up to two-time improvement of the Yb--F distances for the main-cluster atoms. 
Third, in the embedded cluster model one can easily take into account spin-dependent effects with small-core PPs directly and increase basis set size as is done here. 
Fourth, one can apply corrections on accurate wave-function studies (in progress). 
At last, one can consider localized processes, oscillations and waves (electronic excitations, discrete breathers, molecular rotors, etc) which broken crystal symmetry that can not be easily done in periodic studies.

\section{Acknowledgment}

This study was supported by the Russian Science Foundation (Grant No.~14-31-00022). 
We are grateful to I.V.\ Abarenkov and A.V.\ Zaitsevskii for many fruitful discussions.
Calculations in the paper were carried out using resources of the collective usage centre ``Modeling and predicting properties of materials'' at NRC ``Kurchatov Institute'' - PNPI.

%

\end{document}